\documentclass[conference]{IEEEtran}
\usepackage{cite}
\usepackage{amsmath,amssymb,amsfonts}
\usepackage{algorithmic}
\usepackage{graphicx}
\usepackage{textcomp}
\usepackage{ascmac}
\usepackage{xcolor}
\usepackage{lscape}
\def\BibTeX{{\rm B\kern-.05em{\sc i\kern-.025em b}\kern-.08em
    T\kern-.1667em\lower.7ex\hbox{E}\kern-.125emX}}
\begin{document}

\title{Who Made This Copy?\\An Empirical Analysis of Code Clone Authorship}

\author{\IEEEauthorblockN{Reishi Yokomori}
\IEEEauthorblockA{\textit{Dept. Software Engineering} \\
\textit{Nanzan University}\\
Nagoya, Japan\\
yokomori@nanzan-u.ac.jp}
\and
\IEEEauthorblockN{Katsuro Inoue}
\IEEEauthorblockA{\textit{Dept. Software Engineering} \\
\textit{Nanzan University}\\
Nagoya, Japan \\
inoue599@nanzan-u.ac.jp}
}

\maketitle

\begin{abstract}

Code clones are code snippets that are identical or similar to other snippets within the same or different files. They are often created through copy-and-paste practices during development and maintenance activities. Since code clones may require consistent updates and coherent management, they present a challenging issue in software maintenance. Therefore, many studies have been conducted to find various types of clones with accuracy, scalability, or performance. However, the exploration of the nature of code clones has been limited.
Even the fundamental question of whether code snippets in the same clone set were written by the same author or different authors has not been thoroughly investigated. 

In this paper, we investigate the characteristics of code clones with a focus on authorship. 
We analyzed the authorship of code clones at the line-level granularity for Java files in 153 Apache projects stored on GitHub and addressed three research questions.

Based on these research questions, we found that there are a substantial number of clone lines across all projects (an average of 18.5\% for all projects). Furthermore, authors who contribute to many non-clone lines also contribute to many clone lines. Additionally, we found that one-third of clone sets are primarily contributed to by multiple leading authors.

These results confirm our intuitive understanding of clone characteristics, although no previous publications have provided empirical validation data from multiple projects. As the results could assist in designing better clone management techniques,  we will explore the implications of developing an effective clone management tool.

\end{abstract}


\section{Introduction}

A code clone, or simply a clone, is a code snippet that has the same or similar snippet in the same or different files\cite{roy07,inoue21-2}. One of the main causes of code clones is the practice of copying and pasting code during development or maintenance activities. Code clones can create logical dependencies among code snippets and increase the complexity of codebases\cite{lin14}. As a result, they are sometimes considered a bad code smell 
and can have a negative impact on the maintainability of codebases\cite{fowler99, geiger06}.

Research on code clones has been conducted since the 1990s, and since then, a large number of studies have been carried out and published\cite{koschke07,roy07,inoue21,rattan13}. Many of these studies have focused on algorithms, performance (such as recall and precision), and scalability of code clone detection tools, competing with each other\cite{bellon07,roy09, walker19}. Also, there have been empirical studies on code clones, which mainly focus on the evolution of clones and the relation to fault proneness\cite{chatterji10, kim05,tokui20}. These studies have only revealed a small portion of the characteristics of code clones. Knowing when, by whom, in what context, and how code clones were created is important information for managing code clones and supporting developers.
As a first step in understanding the characteristics of code clones, we became interested in investigating the authors of code clones.

Author information is a valuable resource for effective software maintenance\cite{linares12}. However, there is limited knowledge about the authorship of clones. Previous research on clone authorship has primarily focused on a small set of projects, with an emphasis on change proneness and reusability\cite{harder12,moriwaki14}.

In this paper, we conduct an empirical study on the authorship of clones in the context of open-source software (OSS) collections, specifically focusing on 153 Apache projects written in Java on GitHub. Authorship information is obtained using the \texttt{git blame} command with a granularity of line in each Java file.
Our analysis for those target projects mainly focuses on three research questions: (1) basic statistics of clones, (2) author's contributions to clone and non-clone lines, and (3) characteristics of single and multiple author clone sets.

Through the analysis, we found that there are a substantial number of clone lines existing in all projects, and authors who contribute to many non-clone lines also contribute to many clone lines. Additionally, we found that one-third of clone sets are mainly contributed by multiple leading authors.

These findings support some aspects of previous works and our natural understanding of code clones, with empirically validated data from OSS projects. Additionally, they provide more conclusive empirical evidence of the necessity for tool assistance in effective and secure clone management. We will further discuss the implications of these findings.

The major contributions of this paper are twofold. Firstly, the authorship of code clones has been thoroughly investigated across projects, and empirical validation has been provided for the characteristics of code clones. Secondly, we have shown that the presence of code clones in most projects is high, and many of the clones are authored by multiple authors. This highlights the necessity of code clone management tools for the safe and consistent treatment of code clones.

The paper is organized as follows. Section II describes the terms and their definitions used in this paper, and Section III discusses related works. In Section IV, we present the approach of this research with three research questions. The results of the analysis are shown along with the research questions in Section V. Section VI discusses the results and findings, with implications for tool support. In Section VII, the threats to the validity of this study are mentioned. We conclude our discussion with future work in Section VIII.



\section{Terms and Definitions}

A \emph{code snippet}, or simply a \emph{snippet}, is a part of a source code file within a software system. Sometimes, we duplicate a code snippet by copying and pasting it, then modify the pasted part by changing the variable names or literals,  within the same software system or across different ones.

A pair of two code snippets that are the same or similar is called a \emph{clone pair}, and 
each snippet of the pair is referred to as a \emph{code clone} or \emph{clone}\cite{inoue21-2,roy07}. 
In this paper, we use the term `clone' to mean a code snippet that has another code snippet of a clone pair in the same file or different files in the same project. We do not consider here inter-project clone pairs.
The \emph{length} of a clone (or \emph{clone length}) is the lines of code (LOC) of the clone, which may
include non-executable lines such as comments or blank lines.
A \emph{clone set} is a set of code snippets in which any two elements form a clone pair. 
The \emph{size} of a clone set is the number of elements (instances) of the clone set.
The length of a clone set (or clone set length) is the average length of each clone in the clone set.
A \emph{clone line} is a line that is involved in a clone. Conversely,  \emph{non-clone line}
is a line that is not involved in any clone.

A clone pair is generally categorized by its similarity levels as follows\cite{carter1993clone,roy07,inoue21-2}.

\emph{Type-1} clone pair is syntactically identical code snippets, with possible differences of non-executable elements such as white spaces, tabs, comments, and so on.
\emph{Type-2} pair is structurally identical code snippets, with possible differences of identifier names, literals, and type names, in addition to the differences of type-1.
\emph{Type-3} pair is similar code snippets, with possible differences of several statements added, removed, or modified to another snippet, in addition to the differences of type-2. 
\emph{Type-4} pair is syntactically different code snippets, but their functionalities are the same.


    
    
 
    


In this paper, type-1, type-2, and type-3 clones are our targets. Specifically, Type-3 clones are addressed through the composition of type-1 or type-2 clone snippets. Type-4 clones are interesting, but since their possibility of being created by the copy-and-paste actions is low, they were excluded from the scope of this study.

The \emph{author} of a line in a file is the author name given by the \texttt{git blame} command\cite{gitblame22}.
\texttt{git blame} is a command in Git that allows us to see who last modified each line.
It shows each line of the file with a commit hash, the name of the author who last modified the line, and the date and time the modification was made.
In this study, we only use the name of the author.

\begin{figure*}[ht]
\centerline{\includegraphics[width=12.0cm]{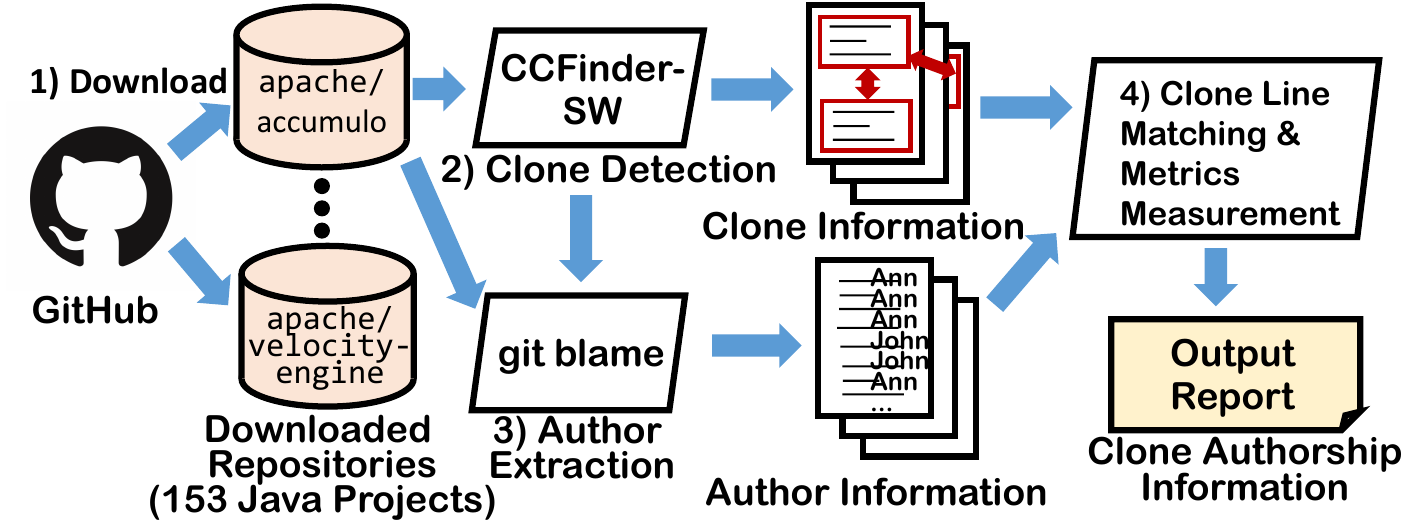}}
\caption{Overview of Analysis Method}
\label{fig:overview}
\end{figure*}

\section{Related Works}

\subsection{Code Clone Analysis}


Numerous studies on code clones have been performed and presented\cite{koschke07,roy07,inoue21,rattan13}. In the 1990s, early research papers were published on detecting methods and fundamental concepts of the code clone\cite{Baker92aprogram,carter1993clone,baxter98}, Since then, many detection methods and tools have been published\cite{Kamiya2002CCFinder,Cordy2011NiCad,Sajnani2016SourcererCC}, and currently, detection methods using machine learning and type-4 clone detection are actively being researched\cite{OREO-FSE2018,Zhao2018DeepSim,arshad22,rabbani22}.

In this study, we use a clone detector to find mainly type-1 and type-2 clones. As type-3 clones are mainly composed of smaller clones of type-1 or type-2, many of their parts are detected and reported in the results of this study.

\subsection{Empirical Studies on Code Clones}


Also, there have been many empirical studies on code clones. Goon et al. examined the ratio of code clones in C and C++ programs in OSS projects over time\cite{goon17}. Honda et al. investigated the change in code clone ratio from the initial development phase to maturity, finding that roughly 20\% of the entire code consists of clones\cite{honda14}. 

Harder investigated code clone authorship and changeability of clones for their clone detection tool as the analysis target\cite{harder12}.
His study found that the clone sets (classes) with a single author accounted for 66.3\% of all clone sets,  and that clone sets with multiple authors were twice as likely to be changed as those with a single author.
While these studies provide specific information on the ratio and the authorship of clones, they do not thoroughly examine the relationship between clones and their authors. Therefore, in this study, we conducted a detailed investigation on code clones and their relationship to their authors.

Moriwaki et al. made an inter-project analysis of OSS projects and identified who reused whose source code across multiple repositories\cite{moriwaki14}. 
Although it provides an interesting perspective on the propagation of code snippets through projects and repositories, no detailed analysis of the authorship of clones inside a single project.

In this paper, we are interested in the details of code clone authorship in the granularity of the source code lines in a clone and code clone set.

\subsection{Code Authorship}

Attributing code authorship is an emerging topic in many areas of software engineering such as code security\cite{kalgutkar19}, plagiarism detection\cite{lee22,dauber19}, and bug triaging\cite{hu14}. These works mainly focus on identifying the authors of code snippets in source or binary forms by using comprehensive methods with the characteristics and statistics of codebases and their archives. 

In this study, we rely on the feature of \texttt{git blame}\cite{gitblame22} to attribute the authorship of code, although it might sometimes produce incorrect information\cite{bird09}. We will discuss the threat of relying
on the \texttt{git blame} command for code authorship attribution in Section \ref{sec:threat}.

\section{Research Approach}

\subsection{Overview}

Fig.\ref{fig:overview} shows an overview of our analysis method, which is described in the following steps.

\begin{enumerate}
    \item First, we have downloaded repositories to analyze from GitHub. The target repositories are mentioned in the following section \ref{sec:target}. The following steps are executed for each repository.

    \item Clone detection is performed by CCFinderSW\cite{semura17}, which is a token-based type-1 and -2 clone detector with a flexible tokenizer that can be adaptable to many languages. CCFinderSW was chosen because it is written only in Java and it works on various environments with high reliability and performance. We ran it with the following parameters: minToken=50 (the minimum number of tokens to detect), rnr=0.5 (the minimum rate of non-repeating parts in the clone fragments), and tks=12 (the minimum number of token types in the clone fragments). These parameters have been chosen to filter out smaller and accidental similar code snippets from the output.
    The output of CCFinderSW is composed of the target file information and the clone set information, both of which are used for the following steps.

    \item \texttt{git blame} commands are executed for all source files to extract author information for each line from the repository. Although we could specify the lines to extract only clone lines using the -L option, for the purpose of measurement and comparison with non-clone lines, we have extracted all line information.

    \item The location of code clones obtained from CCFinderSW and the author information obtained from \texttt{git blame} are matched, and then the author information for the clone lines is obtained. Various statistics are measured and reported as the output.

\end{enumerate}

After completing step 1), the total execution time for steps 2)-4) to analyze the Apache Ant project was 226 seconds on a Ryzen 9-5900HX 32GB Windows 10 machine. Out of this total time, 28 seconds were spent on the clone detection step 2), 193 seconds were used for \texttt{git blame} in step 3), and 5 seconds were spent on clone line matching and metrics measurement in step 4), respectively.

\subsection{Target Selection}
\label{sec:target}

We selected the Apache project's repositories on GitHub as the target of our empirical investigation for the following reasons:

\begin{itemize}
    \item Apache projects are open-source and typically supported by a community of contributors, making them suitable for our empirical investigation. We anticipated that the clones of these repositories would exhibit distinct characteristics compared to the previous research that was mostly developed by a single author.
    
    \item The Apache projects on GitHub encompass a diverse range of repositories, varying in size from small to large with a small to large number of contributors.
    
    \item Many Apache projects host their repositories on GitHub and are easily downloadable for coherent analysis.
\end{itemize}

We have identified and selected 166 Apache-owned projects on GitHub, meeting specific criteria including repository sizes between 100MB and 1GB, more than 100 followers, and written in Java programming language. These constraints were used to match computing resource limitations and to exclude projects that are not popular. 
 13 projects were excluded because they had only a single author, leaving us with 153 projects for the subsequent analysis.\footnote{Note that these projects were downloaded from GitHub between February and March 2023. The complete list of the projects can be found in Table \ref{tab:full-list} in Appendix.}.
To analyze each project, we chose the latest Java files from its project repository, excluding test files that contain the term "test" in their file or path names.


\begin{figure}[t]
\centerline{\includegraphics[width=9.0cm]{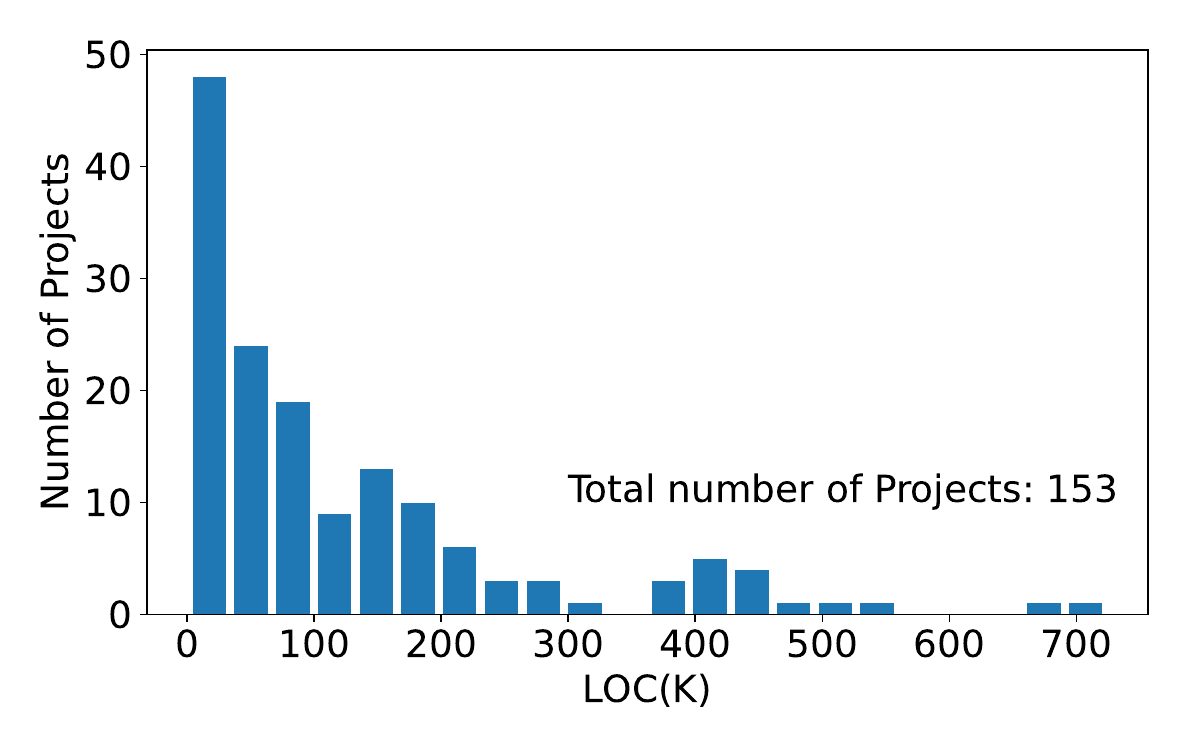}}
\caption{Distribution of the Total Lines of Code in Repositories}
\label{fig:Dist-loc}
\end{figure}

The size of the target Java files, measured in lines of code (LOC), ranges from 1.1K to 723.7K lines, with an average size of 125.9K lines and a median size of 69.2K.
Fig.\ref{fig:Dist-loc} is the distribution of the sizes, which shows that the projects are generally small less than 200K lines. The details of each project size can be seen in Table \ref{tab:full-list} in Appendix.

\subsection{Research Questions}

Firstly, we set a research question (RQ) to investigate the basic information of code clones contained in each project.

%

\begin{screen}[4]
\textbf{RQ1: What is the basic statistical information on code clones of the projects?}
\end{screen}

For each project, we will measure the total lines of code, the total clone lines, the clone length, and the clone set size. We will also calculate the ratio of clone lines to non-clone lines in each project.

Next, we examine the fundamental characteristics of authors who contributed to the code clones.

\begin{screen}[4]
\textbf{RQ2: Do authors who contribute to many non-clone lines also contribute to many clone lines?}
\end{screen}

We will analyze the contribution ratio of each author to clone lines versus non-clone lines and explore the regression between the two.

Thirdly, we examine the authorship of clones within a single clone set.

\begin{screen}[4]
\textbf{RQ3: Are the clones in a clone set contributed by the same author, or
 are they contributed by different authors?}
\end{screen}

We will conduct an analysis of the authorship of clones within clone sets, by identifying clone sets contributed mainly by a single author. Also, we will examine the characteristic differences of clone sets contributed by single authors or multiple authors.



\begin{table}[htbp]
\caption{Basic Statistics of apache/ant}
\begin{center}
\begin{tabular}{|l|r|}
\hline
Repo's name & apache/ant \\
\hline
Repo's size & 97.2MB \\
\hline
Number of target Java files (excluding test files) &  938 \\
\hline
Total lines of target Java files & 235,939 \\
\hline
Clone lines (ratio to total) &   33,064 (\textbf{14.0\%})\\
\hline
Non-clone lines (ratio to total) & 202,875 (86.0\%) \\
\hline
Clone lines Ratio to non-clone lines & 16.3\% \\
\hline
Number of clone set &  \textbf{1,000}\\
\hline
Max, min, average, and median clone length & 260, 1, \textbf{21.8}, 18\\
\hline
Max, min, average, and median clone set size & 36, 2, \textbf{2.67}, 2\\
\hline
Number of different authors in total lines & 87 \\
\hline
Number of different authors in clone lines & 60 \\
\hline
Number of different authors in non-clone lines & 86 \\
\hline
\end{tabular}
\label{tab:ant}
\end{center}
\end{table}

\begin{table}[htbp]
\caption{Basic Statistics of All 153 Projects}
\begin{center}
\begin{tabular}{|l|r|}
\hline
Repositories (153 in total) & apache/accumulo \verb|~| \\
   & apache/velocity-engine \\
\hline
Repo's size & 0.66- 3660 (74.8) MB \\
\hline
Number of target Java files (excl. test files) &  7-3770 (791.5) \\
\hline
Total lines of target Java files & 1.1-723.7 (125.9) K\\
\hline
Clone line ratio to total & \textbf{4.1-64.1 (18.5)} \%\\
\hline
Non-clone line ratio to total & 35.9-95.9 (81.5) \%\\
\hline
Clone lines ratio to non-clone lines & \textbf{4.2-178 (24.6)} \% \\
\hline
Number of clone set &  4-11358 (1274) \\
\hline
Average clone length & 9.1-45.7 (17.3) \\
\hline
Average clone set size & 2.1-15.2 (3.3)\\
\hline
Num. of different authors in total lines & 3-368 (56.0) \\
\hline
Num. of different authors in clone lines & 2-225 (34.5)\\
\hline
Num. of different authors in non-clone lines & 3-361 (55.0)\\
\hline
\end{tabular}
\rightline{(The values inside parentheses are the average for all projects)}
\label{tab:all-proj}
\end{center}
\end{table}

\section{Result}
\label{sec:result}

As an example of project analysis, we will first present the results of the Apache Ant project, followed by the analysis of the set of all projects.

\subsection{RQ1: What is the basic statistical information on code
clones of the projects?}

\begin{figure}[htbp]
\begin{center}
\includegraphics[width=8.8cm]{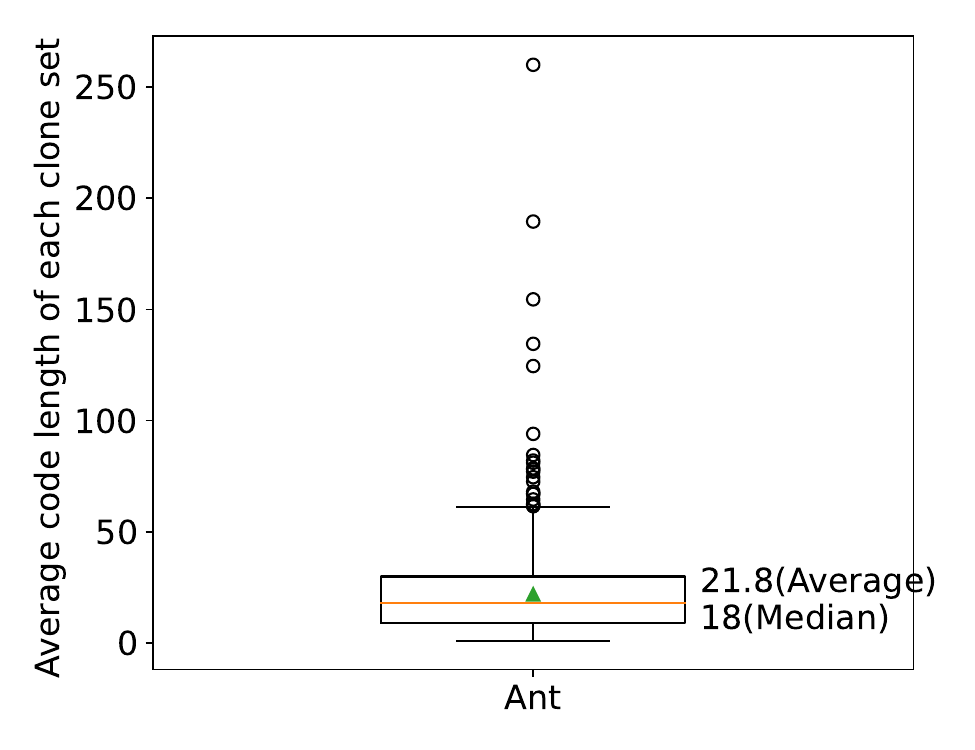}
\caption{Distribution of Ant's Clone Length (LOC)}
\label{fig:ant-clone-length}
\end{center}
\end{figure}

\begin{figure}[htbp]
\includegraphics[width=8.8cm]{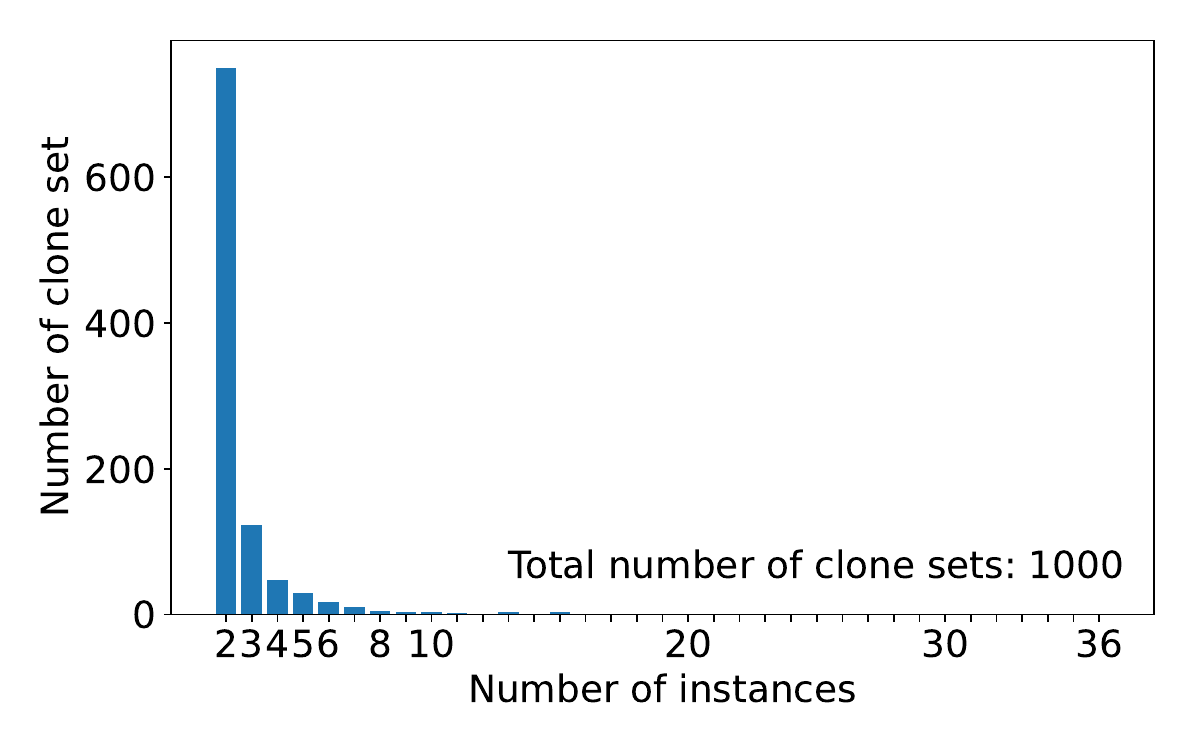}
\caption{Distribution of Ant's Clone Set Size}
\label{fig:ant-clone-size}
\end{figure}

\textbf{(Ant Project)} ~~The basic statistics for Apache Ant and its clones are presented in Table \ref{tab:ant}. As shown in the table, the ratio of clones to the total lines of the target Java files is 14.0\%, and the number of clone sets is 1,000.
Fig.\ref{fig:ant-clone-length} shows in a boxplot the distribution of clone length for each clone set in Ant, and we can see many small clones with some exceptions over 250 lines. The average clone length is 21.8 lines.
As Fig.\ref{fig:ant-clone-size} shows, the distribution of the clone set sizes (the number of instances of clone set) indicates that the majority of sets have a size of 2, with an average size of 2.67.


\textbf{(All Projects)} ~~We extend the basic statistics to all the projects as summarized in Table \ref{tab:all-proj}, 
and some of those metrics are presented in Table \ref{tab:full-list} in Appendix.
Fig.\ref{fig:all-3box}(left) presents a box plot of the distribution of the clone line ratio to the total, 
ranging from 4.1-64.1\% and the average of \textbf{18.5\%}, which is a substantial ratio to the total lines.
Fig.\ref{fig:all-3box}(middle) shows the average length of clones and Fig.\ref{fig:all-3box}(right) shows the size of clone sets for each project. As we can see from these, the average clone length is \textbf{17.3} and the average size is \textbf{3.33} for all projects. This result shows that the clone length is not so small and that more than two copies exist in general.

\begin{figure}[htbp]
\includegraphics[width=8.8cm]{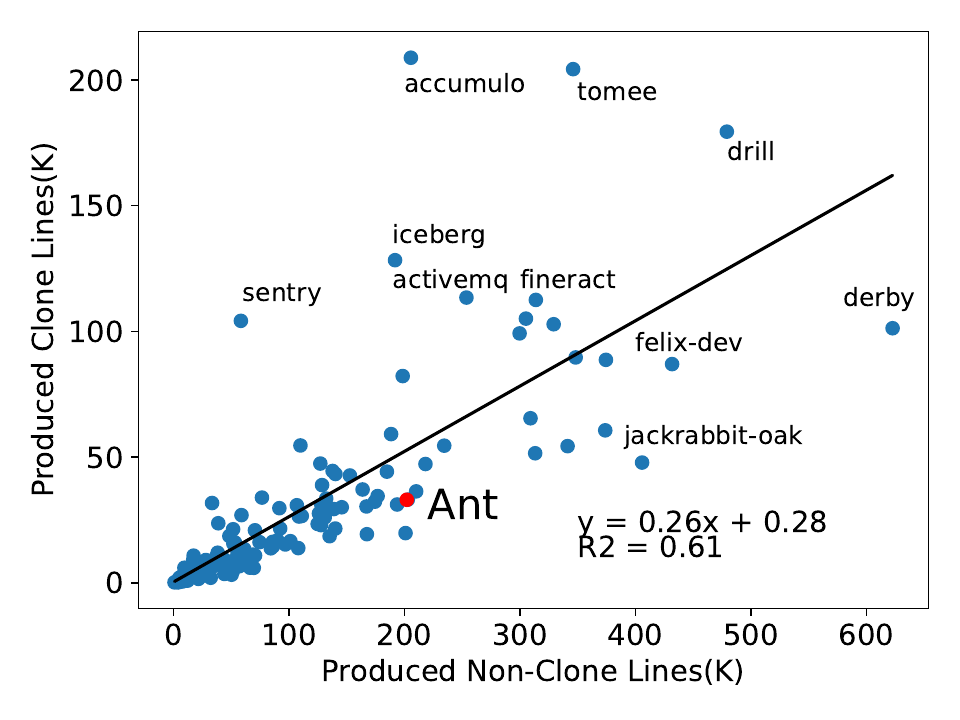}
\caption{Correlation of the Number of Non-Clone Lines vs Clone Lines for All Projects}
\label{fig:inter-plot-cl-ratio}
\end{figure}

We calculated for all projects the ratio of clone lines to non-clone lines and found that it ranged from a minimum of 4.2\% to a maximum of 178\%, with an average ratio of \textbf{24.6\%}. We plotted each project with the number of non-clone lines on the x-axis and clone lines on the y-axis in Fig.\ref{fig:inter-plot-cl-ratio}. The analysis yielded a linear regression coefficient of 0.261 with $R^2$ of 0.615 and a p-value of 2.54E-33, demonstrating the correlation coefficient's significance. 

\begin{screen}[4]
\textbf{Answer to RQ1: Substantial number of clone lines exist in all projects. Averagely 18.5\%  of all lines and 24.6\% of non-clone lines  are clone lines, with average clone lengths of 17.3 lines and average clone set sizes of 3.33.
There is a statistically significant correlation between the number of clone lines and the number of non-clone lines.
}
\end{screen}


\begin{figure}[htbp]
\includegraphics[width=2.8cm]{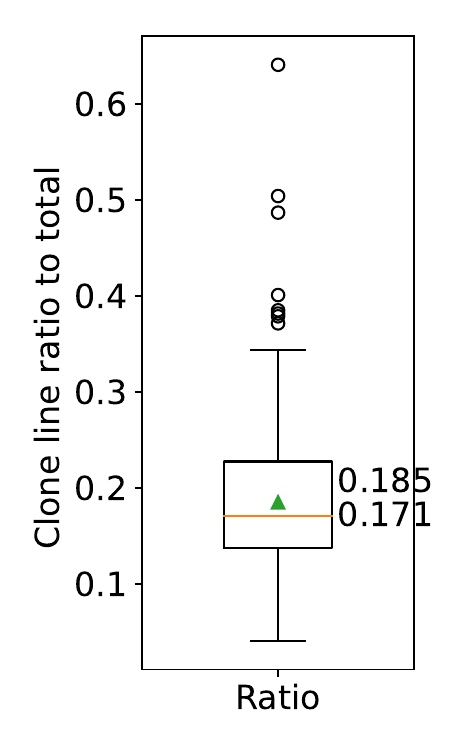}
\includegraphics[width=2.8cm]{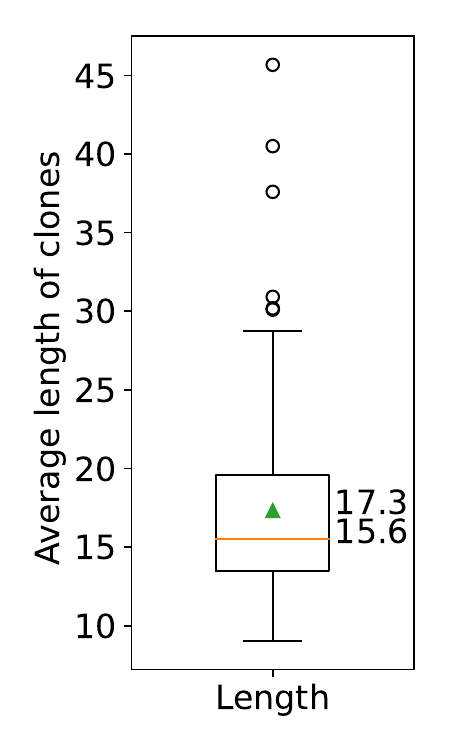}
\includegraphics[width=2.8cm]{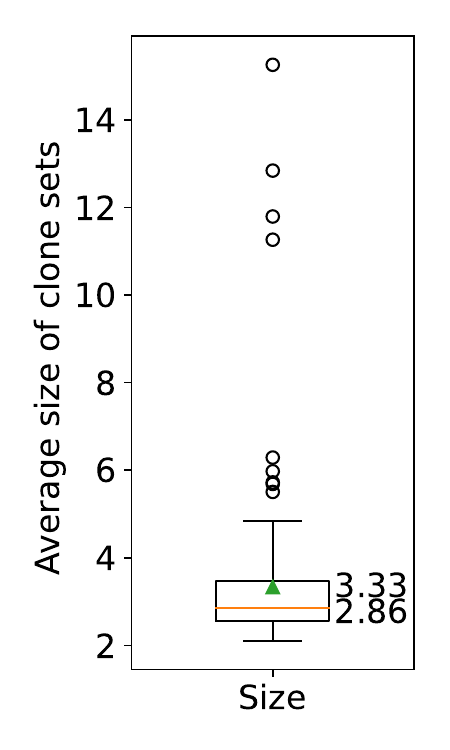}
\caption{Clone Line Ratio (left), Clone Line Length (middle), and Clone Set Size (right) for All Projects}
\label{fig:all-3box}
\end{figure}

\subsection{RQ2: Do authors who contribute to many non-clone lines also contribute to many clone lines?}

\begin{figure}[htbp]
\begin{center}
\includegraphics[width=8.0cm]{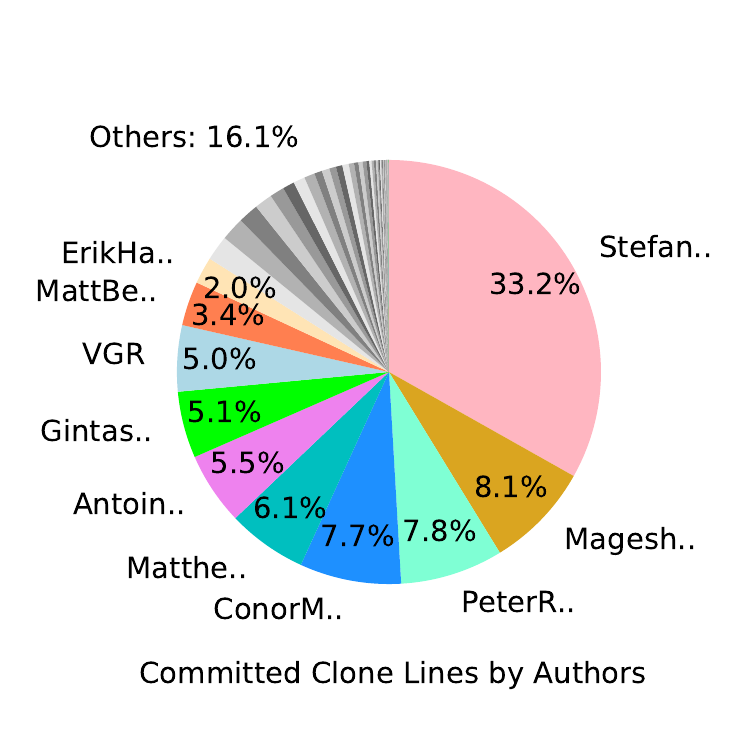}

\includegraphics[width=8.0cm]{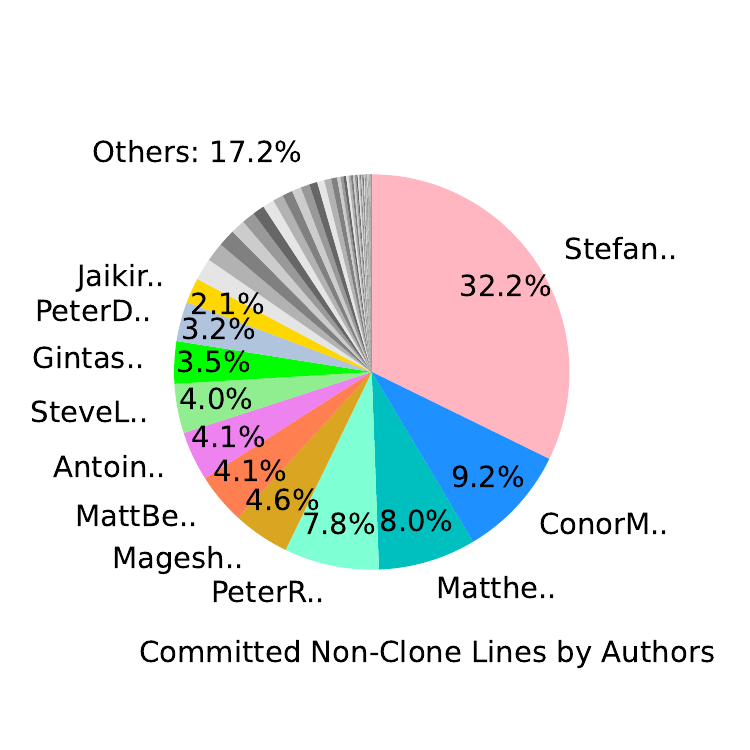}
\caption{Ratio of Contributed Clone and Non-Clone Lines by Authors}
\label{fig:auth-cl-noncl}
\end{center}

\end{figure}

\textbf{(Ant Project)} ~~In Ant project, there are 87 different authors in the total lines, with 86 in the non-clone lines and 59 in the clone lines in the target Java files.
Fig.\ref{fig:auth-cl-noncl} shows the ratio of the contributed clone lines (upper figure) and non-clone lines (lower figure).

\begin{figure}[htbp]
\includegraphics[width=8.8cm]{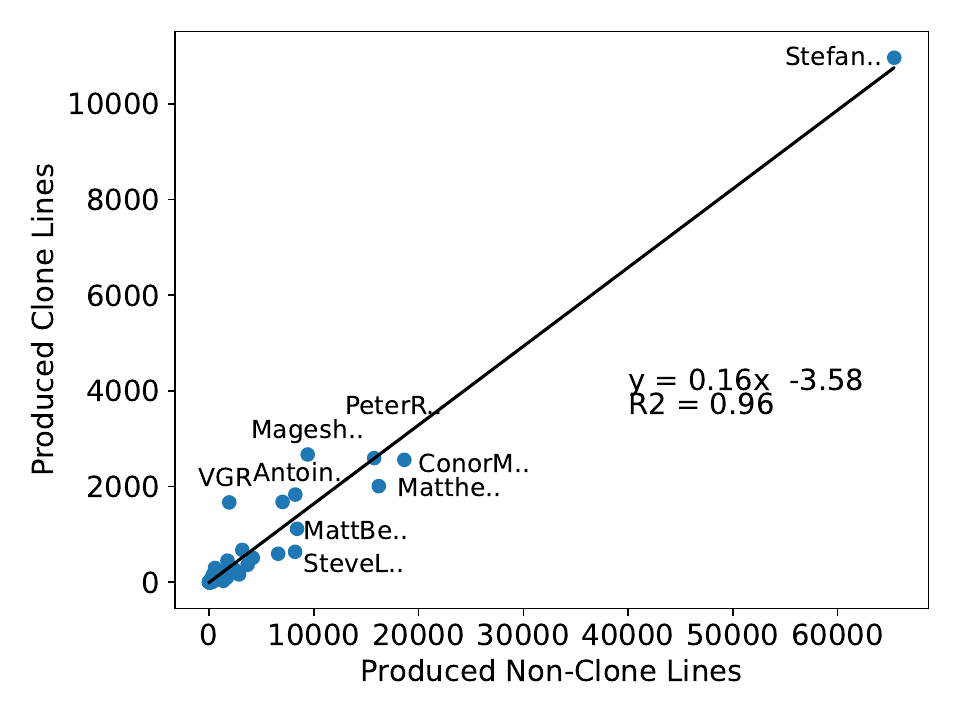}
\caption{Correlation of Author Contribution to Clone and Non-Clone Lines in Ant}
\label{fig:regression}
\end{figure}

As can be seen from these figures, although there are some differences in the order of contribution ratio, the distribution of contributing amounts by the authors is similar. This suggests that the contributions of non-clone line and clone line authors are strongly related. To quantitatively validate this, we examined the correlation between the amount of author contribution to non-clone lines and clone lines, as shown in Fig.\ref{fig:regression} where each dot represents one author with the number of contributed non-clone lines on the x-axis and that of contributed clone lines on the y-axis. We determined a linear regression coefficient of 0.164 with a coefficient of determination $R^2$ of 0.96 and a P-value in null-hypothesis significance testing of 0.0. These values show the statistically significant correlation between the two, and the regression coefficient of 0.164, which indicates that authors are contributing to the creation of clone lines that account for 16.4\% of non-clone lines.

\textbf{(All Projects)} ~~We also tested the significance of the correlation coefficient between non-clone and clone lines contributed by the same authors in all 153 projects (see details in Appendix Table \ref{tab:full-list}). Our results showed that, while three projects were excluded, the correlation coefficient was significant in \textbf{150 projects}.

\begin{screen}[4]
\textbf{Answer to RQ2: Yes, authors who contribute to many non-clone lines also contribute to many clone lines in 150 out of 153 projects.}
\end{screen}

\subsection{RQ3: Are the clones in a clone set contributed by the same author, or
 are they contributed by different authors?}

To perform this analysis effectively, we define two types of clone sets: \emph{single-leader} clone set and \emph{multi-leader} clone set.
For a code snippet $s$, the \emph{leading author} or simply \emph{leader} $l_s$ of $s$ refers to the author who has committed the most lines in $s$, i.e., the top contributor in the snippet\footnote{Note that if there are two or more such authors with the same number of committed lines, the leader is randomly selected from them.}.
For a clone set $C$ with clones $c_1$, $c_2$, \ldots, if all leaders of  $c_1$, $c_2$, \ldots, are the same, then we say $C$ is a single-leader clone set. Otherwise, it is called a multi-leader one.

The leader is the most contributing author. A single-leader clone set means that the leaders of each clone in the clone set are the same, even if there are different authors for a small portion of the clones.  The reason for using leader to investigate the authorship of clones is that we want to know an overview of the dominant authors of clones, without being affected by minor contributions. 

\begin{figure}[htbp]  
\includegraphics[width=8.8cm]{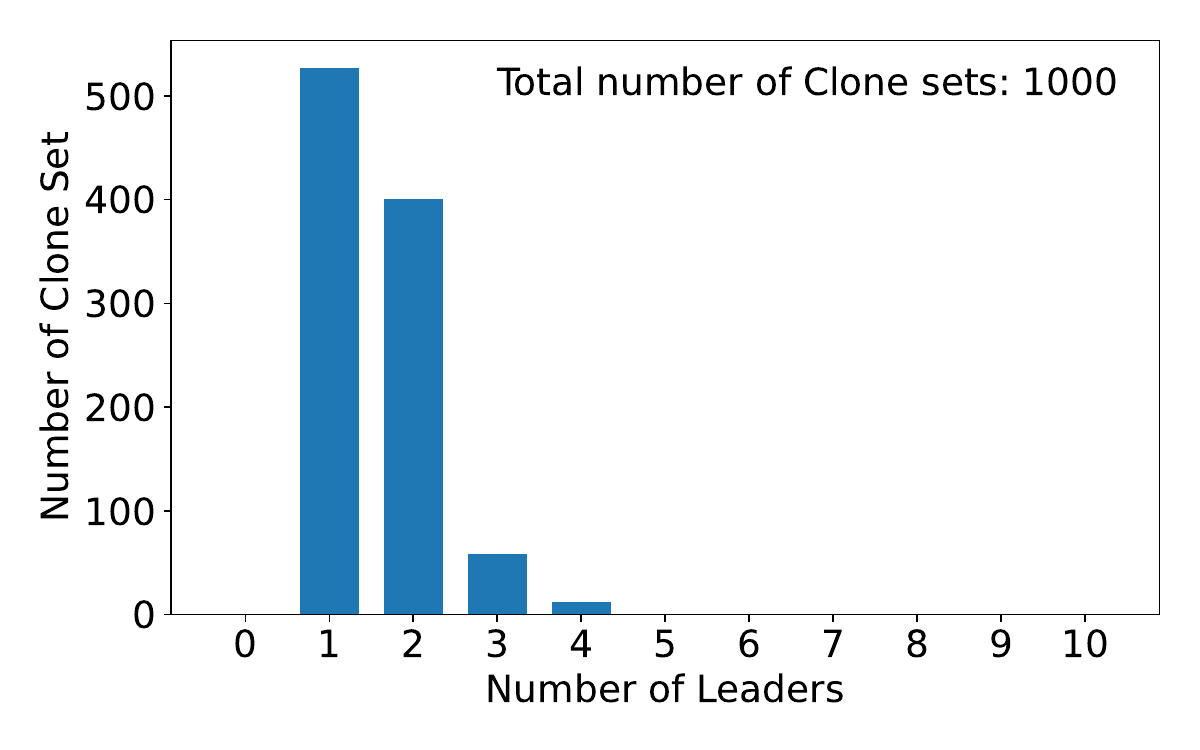}
\caption{Number of Leaders in Clone Set in Ant}
\label{fig:ant-leader}
\end{figure}

\textbf{(Ant Project)} ~~Firstly, we analyze the proportion of single-leader and multi-leader clone sets in the Ant project. Out of exactly 1,000 clone sets we have found in the project, 527 (52.7\%) are classified as single-leader, while 473 (47.3\%) are multi-leader. Therefore, it can be concluded that approximately half of the clone sets in the project have multiple leaders. Fig.\ref{fig:ant-leader} shows the breakdown of the  number of leaders in each clone set, and it is clear that most multi-leader clone sets have two leaders.

\begin{figure}[htbp] 
\begin{center}
\includegraphics[width=8.0cm]{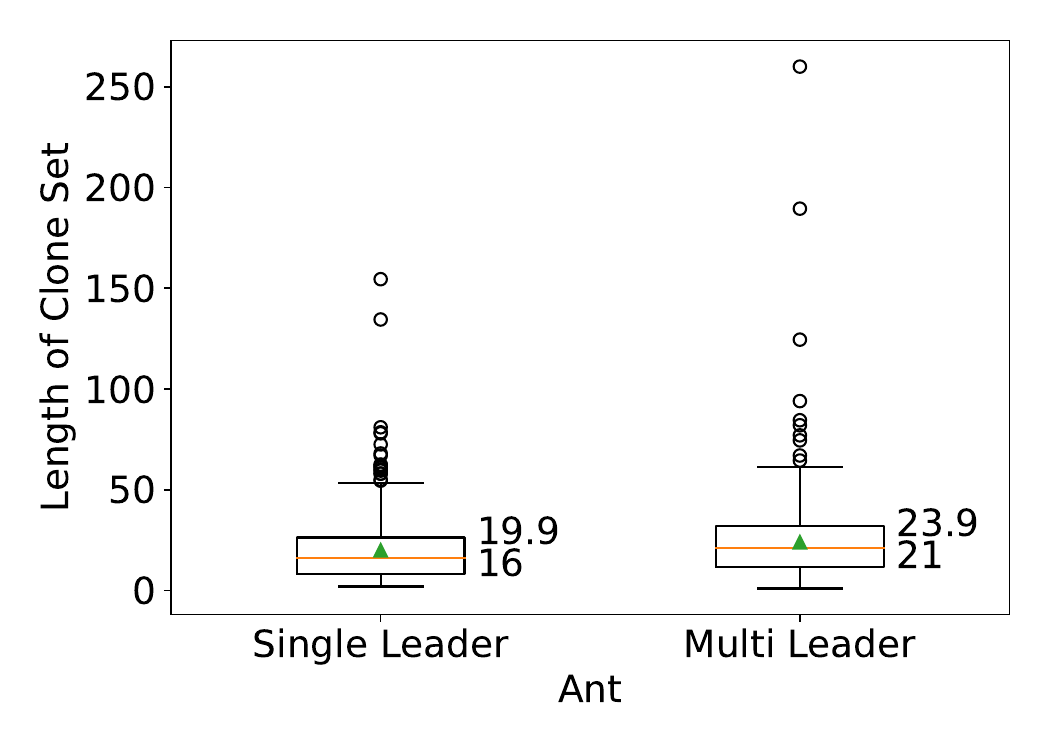}
\caption{Length of Single and Multi-Leader Clone Sets in Ant}
\label{fig:ant-length-s-m}
\end{center}
\end{figure}

\begin{figure}[htbp]  
\begin{center}
\includegraphics[width=8.0cm]{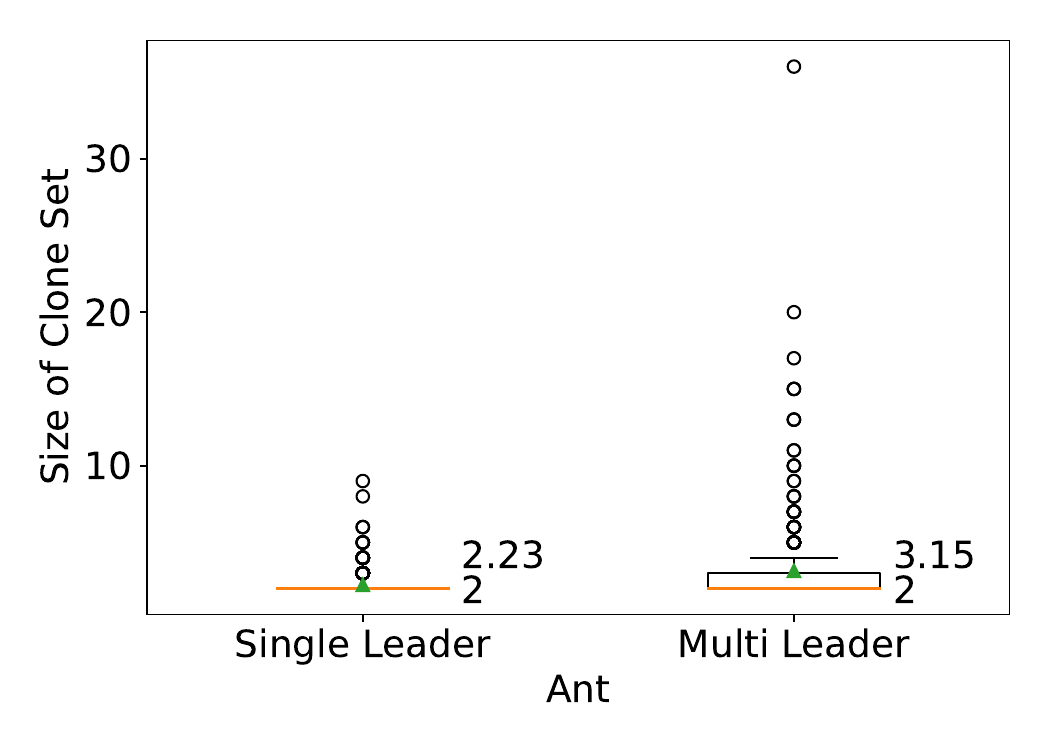}
\caption{Size of Single-and Multi-Leader Clone Sets in Ant}
\label{fig:ant-size-s-m}
\end{center}
\end{figure}

Next, we conducted an investigation into the characteristics of single- and multi-leader clone sets in the Ant project. The lengths of the single- and multi-leader clone sets were measured and depicted as box plots in Fig.\ref{fig:ant-length-s-m}. The average lengths of the single- and multi-leader clone sets were found to be 19.9 and 23.9 lines, respectively, with median values of 16 and 21 lines, respectively. The Mann-Whitney U-test revealed a significant difference in clone length ($p=2.14E-06$) between the two sets, indicating that clones in multi-leader clone sets tend to be longer than those in single-leader sets.

Similarly, we examined the sizes of single- and multi-leader clone sets, as illustrated in Fig.\ref{fig:ant-size-s-m}. The average sizes of the single- and multi-leader clone sets were found to be 2.23 and 3.15, respectively, with both having median values of 2. The Mann-Whitney U-test showed a significant difference ($p=4.89E-17$) between the two sets, indicating that multi-leader clone sets tend to be larger in size than single-leader sets.

\textbf{(All Projects)} ~~
We have extended the analysis to all projects.
The ratio of single-leader clone sets to all clone sets ranges from 17.3\% to 99.2\%, with both an average and median of \textbf{66.7\%} (see the details in Table \ref{tab:full-list} in Appendix). This means that the remaining \textbf{33.3\%} of clone sets are multi-leader and composed of different leader authors.

We had expected that the ratio of the multi-leader clone sets to all clone sets, ranging from 0.8\% to 82.7\%, would correlate to the size of the project (total lines of code) or the total number of authors. However, we found no correlation between them. This means that \textbf{an increase in total lines or committed authors does not necessarily lead to an increase in multi-leader clone sets}.

We have examined the difference in clone lengths between single- and multi-leader clone sets for all projects, as we did for the Ant project. The Mann-Whitney U-test was performed independently for each project to determine if there was a significant difference in clone lengths, with an average longer multi-leader clone set than the single one. Our analysis confirmed that out of all 153 projects, 34 had longer multi-leader clone sets than single-leader sets. For the remaining 119 projects, which comprise the majority, there was no significant difference in the lengths of the single- and multi-leader clone sets.

We performed the Mann-Whitney U-test for each project to determine if there was a significant difference in the clone set sizes between single- and multi-leader sets. Our analysis revealed that 105 out of 153 projects (\textbf{68.6\%}) showed larger sizes of the multi-leader clone sets with significant differences.  This means that in two-thirds of the projects, the sizes of the multi-leader clone sets, which is the number of instances of each clone set,  are larger than those of the single-leader sets.

\begin{screen}[4]
\textbf{Answer to RQ3: One-third (33.3\%) of the clone sets across all projects are contributed by multiple leaders, while the remaining two-thirds (66.7\%) are contributed by a single leader. An increase in total lines or committed authors does not necessarily lead to an increase in multi-leader clone sets. In approximately two-thirds (68.6\%) of all projects, multi-leader clone sets have larger clone set sizes compared to single-leader ones.}
\end{screen}

\begin{figure}[htbp] 
\includegraphics[width=8.8cm]{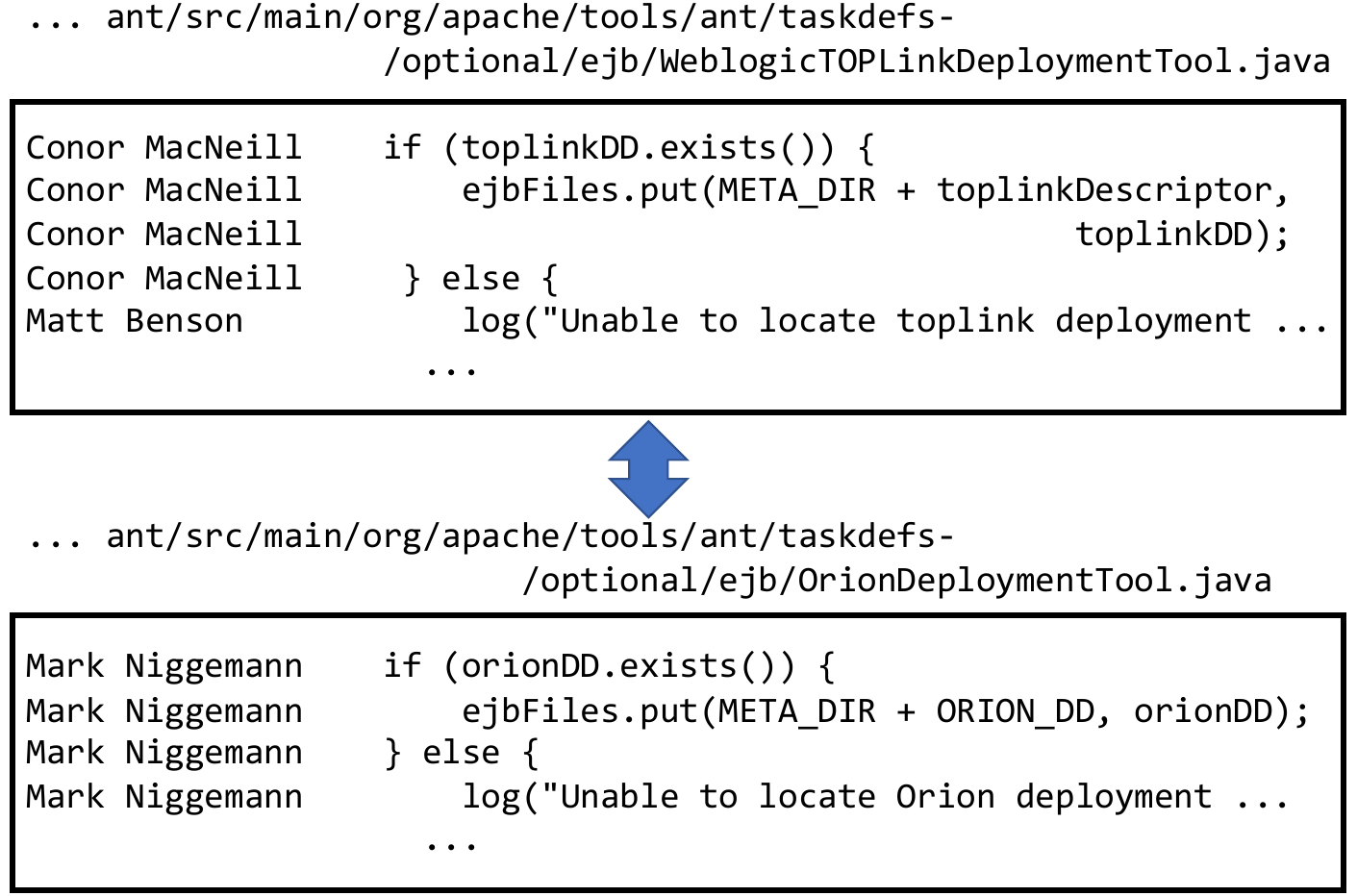}
\caption{An Example of Multi-Leader Clone Set}
\label{fig:ant-code1}
\end{figure}

\section{Discussions}

\subsection{Ratio of clone lines (RQ1)}

Honda et.al have investigated the convergence of code clone ratio in five open-source projects including Firefox and Linux kernel\cite{honda14}. They have found that at first, the ratio varies widely, but as development progresses, it settles at around 20\% for all projects.
In our analysis, the average clone line ratio for all projects is 18.5\%, similar to their work.

Another work for the clone ratio can be found in \cite{goon17}. In this paper, the authors analyzed libcurl, skynet, and git as targets, and found clone line ratios ranging from about 4\% to 9\% of the total lines of code. 
The distinction from our ratio might be due to differences in the targets and the clone detector as well as their measurement method that excludes all comment and whitespace lines in their approach.

\subsection{Contribution to Clone Lines (RQ2)}

The answer to RQ2 states that authors who contributed non-clone lines also contributed clone lines. 
This result aligns very well with our natural intuition for code clone authors
and it is statistically validated with 98\% projects (150 out of 153 projects).
The outlier projects might have authors 
who create mainly code clones but no other code. This might happen if there are developers who mainly perform debugging or maintenance tasks.

\subsection{Different Authors in a Clone Set (RQ3)}
\label{sec:different}
The answer to RQ3 states that two-thirds of clone sets are mainly contributed by single leading authors, while the remaining one-third are mainly contributed by multiple leading authors. Fig.\ref{fig:ant-code1} depicts a case of the multiple leading authors found in the Ant project. In this example, a type-2 clone pair is presented, with the left column showing the authors of the lines in the right column. By the analysis of timestamps, the upper snippet, which was mainly contributed by "Conor MacNeill," is earlier than that of the lower snippet contributed by "Mark Niggemann". It appears that the upper snippet or another clone snippet was copied and pasted to create the lower one. 

It is important to recognize a case such that an author $A$ creates a code snippet $S_A$ earlier and a different author $B$ later copies and makes a snippet $S_B$ so that a code clone pair {$S_A$, $S_B$} is formed. In this case, author $A$ might modify $S_A$ to a new snippet $S_A'$ without knowing the existence of $S_B$, and inconsistency between $S_A'$ and $S_B$ would easily happen. We will discuss this issue later in \ref{sec:implecation}.

\subsection{Multi-Leader Clone Set (RQ3)}
As stated in the answer to RQ3, we employed a method called single- and multi-leader to measure the major contributing authors of a clone set. This method helps to understand the majority of the clone's contributors, suppressing small contributor effects. We also measured the number of clone sets committed by a purely single author, without using the concept of the leader, and calculated the ratio of single-author clone sets to all clone sets. Our results showed that the average ratio of the purely single-author clone set for all projects was 34.4\%, which is different from the 66.7\% of the single-leader clone set ratio. This indicates the presence of minor contributions from non-leading contributors that have affected the ratio. 

In a previous study, the ratio of purely single-author clone sets (classes) was reported as 66.3\%\cite{harder12}. This value is higher than our average of 34.4\%, which can be attributed to the fact that the previous study is an analysis result of a closed development project with limited members, whereas our study analyzed open development projects with the participation of many developers.

\subsection{Implecation for Code Clone Management System}
\label{sec:implecation}

RQ1 shows that approximately 20\% of the source code is composed of code clones, which is consistent with previous research findings \cite{honda14}. Also, RQ2 suggests that no special but ordinary developers or maintainers would easily create, delete, or change code clones during their daily activities. These facts indicate that code clones have a fairly large presence in the source code. As code clones establish logical dependencies between code snippets, special attention must be given to them. Previous research has addressed this topic, suggesting features to identify, track, and report code clones and their changes, with a focus on possible bug-inducing activities\cite{ekoko08,mondal21,tokui20}. Therefore, we consider that source code management tools must have features that can effectively search, trace, and report these logical dependencies generated by code clones.

In addition, a monitoring feature of the clone's consistency, which continuously watches addition, deletion, and changes to all source code, and identifies changes in code clone sets, is important. As mentioned in Sec.\ref{sec:different}, the author of the original snippet might not know the existence of the clones created by other authors, the continuous monitoring by tools would help to happen inconsistency between logically aligned code snippets.

Based on RQ3, one-third of code clone sets are contributed by multiple leaders, i.e., multiple authors. It is intuitive to assume that multiple-author code snippets require more attention than single-author snippets because inconsistent editing may easily occur due to the involvement of multiple individuals. While validating the impact on the code quality of single versus multiple authors on code snippets would be an interesting future research topic, a bug prediction system that incorporates clone information associated with their authorship could expedite software maintenance tasks.

\section{Threat to Validity}
\label{sec:threat}

\noindent
[Internal Validity]

We have used a clone detector CCFinderSW in our study. As reported by other researchers\cite{bellon07, roy09}, different code clone detectors may report clones differently for the same target, which could potentially affect our results. Although the accuracy of CCFinderSW has been confirmed by its developer\cite{semura17}, it would be worthwhile to explore the use of other detectors such as NiCad\cite{Cordy2011NiCad}, CCFinderX\cite{kamiya21}, or SourcererCC\cite{Sajnani2016SourcererCC} to strengthen our result. NiCad or SourcererCC, for example, are capable of detecting type-3 clones directly, which could expand our analysis to include type-3 clones, even though many parts of those could be detected and included as type 1 or 2 clones by our approach.

We have used a default setting of 50 tokens for the 'minToken' parameter of CCFinderSW, which is the minimum token length for detection. This corresponds to a few lines of code in general and may lead to the identification and filtering-out of unintentional and not copy-and-paste token sequences as clones. Additionally, we have also used other options such as 'rnr=0.5' (the minimum rate of non-repeating parts in the clone fragments) and 'tks=12' (the minimum number of token types in the clone fragments) to filter out accidental simple token sequences from the output and mitigate the risk of including accidentally similar snippets. We determined these parameters from default values and simple checks,
but exploring how the results change with various combinations of parameters is future work.

We have relied on the author's information provided by Git for this analysis. However, Git's author information may not always be reliable\cite{bird09}. For example, if the original author used multiple names, they may be recognized as different individuals. Additionally, author information might be overridden by Git commands by intention or accidentally. To mitigate these risks, a potential future research topic would be to employ email addresses or other associated information to more precisely identify authors.

\vspace{3mm}
\noindent
[External Validity]

In this research, we analyzed 153 repositories from Apache projects, which might be a smaller size sample compared to the gigantic real-world OSS projects. However, our sample includes repositories of various sizes with a range of clones, as discussed in Section \ref{sec:target} and \ref{sec:result}. As such, we believe that our findings could be applicable to other OSS projects or software products, although further research would be needed to validate this.

Our analysis focused only on Java programs. Different programming languages may exhibit different characteristics in terms of code clones. Exploring language differences and the characteristics of clones in other programming languages would be an interesting direction for future research.

\section{Conclusion}

We have presented our analysis of 153 Apache projects aimed at characterizing the authors of code clones within these projects. Our findings show that, on average, 18.5\% of the lines of code in all projects are in clones, with an average clone length of 17.3 lines and an average clone set size of 3.33. This indicates a fairly large presence of clone lines in OSS projects, which need to be handled with care during development and maintenance tasks.
Furthermore, our validation has revealed a strong correlation between the contributions of non-clone lines and clone lines, meaning that authors who contribute to many non-clone lines also tend to contribute to many clone lines.
In addition, we have discovered that two-thirds of clone sets are contributed by single leaders, while the remaining are contributed by multiple leaders which might introduce potential inconsistency risks.
These results confirmed our intuitive understanding of code clone characteristics with empirical evidence, and we have explored the implications of developing an effective clone management tool.

In our future work, we plan to continue our empirical investigation of clone authorship in relation to code quality and faults. Previous research has shown that clones with multiple authors are more frequently changed \cite{harder12}. However, the analysis in their study was limited to only one project, and no relation to faults was shown. Therefore, we aim to extend our analysis to multiple target projects and specifically investigate the relationship between clone authorship and the occurrence of faults.

\section*{Acknowledgment}

 This work is supported by JSPS Grants-in-Aid for Scientific Research, Category (B), 23H03375, and Nanzan University Pache Research Subsidy I-A-2 for the 2023 academic year. 
 We are grateful to Ayu Kojima and Yuka Sato for their contributions to the clone data analysis.


\section*{Appendix}

Following Table \ref{tab:full-list} lists all 153 projects used for our analysis, and shows some of the analysis results. One project is composed of 5 lines:

\begin{itemize}
    \item Projects: Project name
    \item LOC(K): The number of K lines analyzed in the project
    \item CL/ALL(LOC): The ratio of the clone lines to all lines
    \item SCS/ALLCS: The ratio of the number of single-leader clone sets to the number of all clone sets
    \item Tests: The following symbols indicate that each significance test is validated, respectively.
        \begin{itemize} 
            \item[*] Significance test of the correlation coefficient between non-clone and clone lines 
            \item[+] Significance difference of the clone lengths between single- and multi-leader clone sets and the multi-leader clone sets are longer than the single ones
            \item[@] Significance difference of the clone set sizes between single- and multi-leader clone sets and multi-leader clone sets are larger than the single ones.
        \end{itemize}
\end{itemize}

\begin{landscape}
\begin{table}[ht]
 \caption{List of Target Apache Projects( 153 Projects)}
  \includegraphics[width=23.5cm]{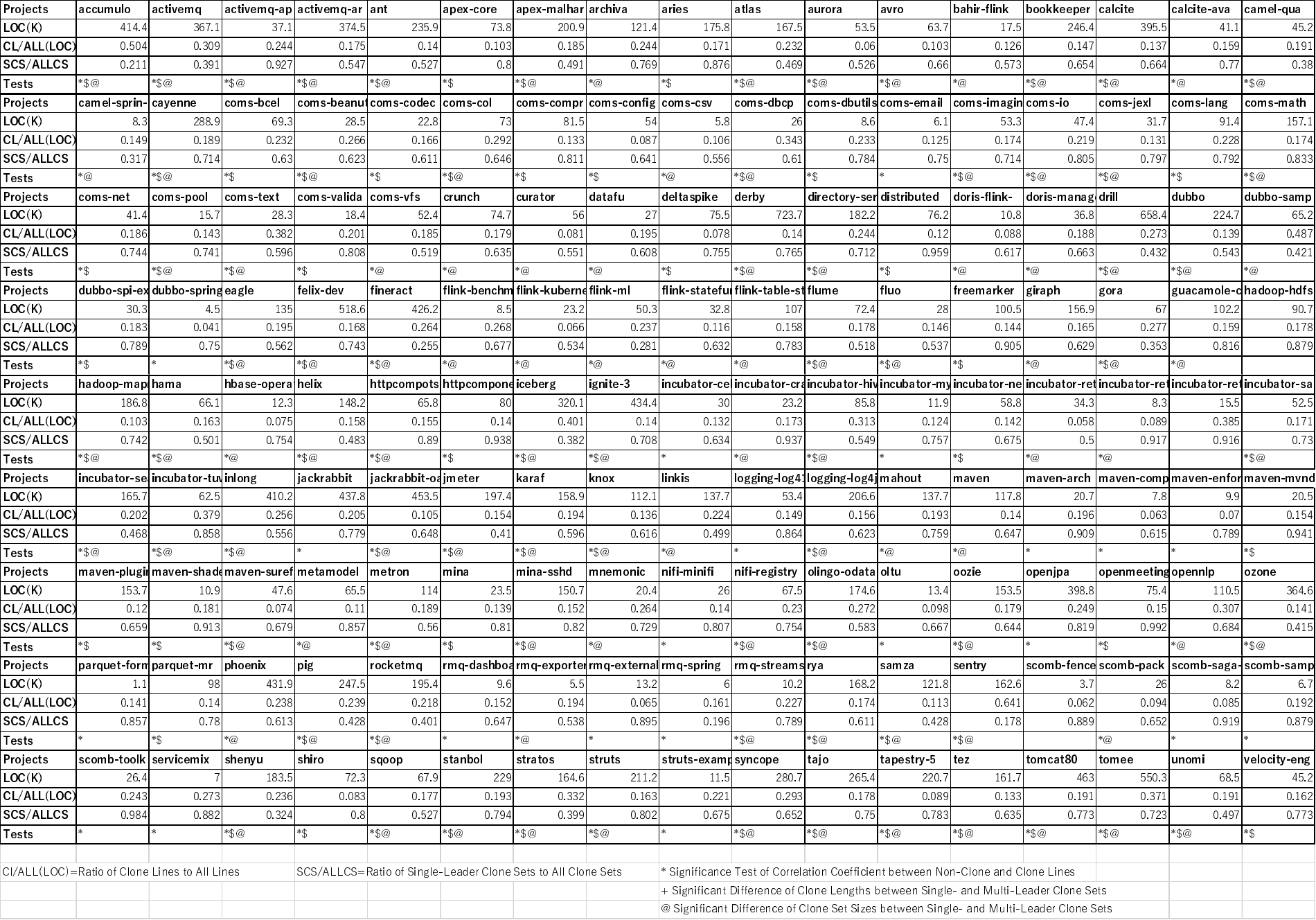}
\label{tab:full-list}
\end{table}
\end{landscape}

\bibliographystyle{IEEEtranS}

\bibliography{cloneAuthor-bib}

\begin{thebibliography}{10}
\providecommand{\url}[1]{#1}
\csname url@samestyle\endcsname
\providecommand{\newblock}{\relax}
\providecommand{\bibinfo}[2]{#2}
\providecommand{\BIBentrySTDinterwordspacing}{\spaceskip=0pt\relax}
\providecommand{\BIBentryALTinterwordstretchfactor}{4}
\providecommand{\BIBentryALTinterwordspacing}{\spaceskip=\fontdimen2\font plus
\BIBentryALTinterwordstretchfactor\fontdimen3\font minus
  \fontdimen4\font\relax}
\providecommand{\BIBforeignlanguage}[2]{{%
\expandafter\ifx\csname l@#1\endcsname\relax
\typeout{** WARNING: IEEEtranS.bst: No hyphenation pattern has been}%
\typeout{** loaded for the language `#1'. Using the pattern for}%
\typeout{** the default language instead.}%
\else
\language=\csname l@#1\endcsname
\fi
#2}}
\providecommand{\BIBdecl}{\relax}
\BIBdecl

\bibitem{arshad22}
S.~Arshad, S.~Abid, and S.~Shamail, ``Codebert for code clone detection: A
  replication study,'' in \emph{2022 IEEE 16th International Workshop on
  Software Clones (IWSC)}, 2022, pp. 39--45.

\bibitem{Baker92aprogram}
B.~S. Baker, ``A program for identifying duplicated code,'' \emph{Proc. of
  Computing Science and Statistics: 24th Symposium on the Interface 24}, pp.
  49--57, 1992.

\bibitem{baxter98}
I.~Baxter, A.~Yahin, L.~de~Moura, M.~Sant'Anna, and L.~Bier, ``Clone detection
  using abstract syntax trees,'' in \emph{Proc. of International Conference on
  Software Maintenance}, vol. 368-377, 01 1998, pp. 368--377.

\bibitem{bellon07}
S.~Bellon, R.~Koschke, G.~Antoniol, J.~Krinke, and E.~Merlo, ``Comparison and
  evaluation of clone detection tools,'' \emph{IEEE Transactions on Software
  Engineering}, vol.~33, no.~9, pp. 577--591, 2007.

\bibitem{bird09}
C.~Bird, P.~C. Rigby, E.~T. Barr, D.~J. Hamilton, D.~M. German, and P.~Devanbu,
  ``The promises and perils of mining git,'' in \emph{2009 6th IEEE
  International Working Conference on Mining Software Repositories}, 2009, pp.
  1--10.

\bibitem{carter1993clone}
S.~Carter, R.~Frank, and D.~Tansley, ``Clone detection in telecommunications
  software systems: A neural net approach,'' in \emph{Proc. Int. Workshop on
  Application of Neural Networks to Telecommunications}, 1993, pp. 273--287.

\bibitem{chatterji10}
D.~Chatterji, B.~Massengill, J.~Oslin, J.~C. Carver, and N.~A. Kraft,
  ``Measuring the efficacy of code clone information: An empirical study,'' in
  \emph{Evaluation and Usability of Programming Languages and Tools}, ser.
  PLATEAU '10.\hskip 1em plus 0.5em minus 0.4em\relax New York, NY, USA:
  Association for Computing Machinery, 2010.

\bibitem{Cordy2011NiCad}
J.~R. {Cordy} and C.~K. {Roy}, ``The nicad clone detector,'' in \emph{2011 IEEE
  19th International Conference on Program Comprehension}, June 2011, pp.
  219--220.

\bibitem{dauber19}
E.~Dauber, A.~Caliskan, R.~Harang, G.~Shearer, M.~Weisman, F.~Nelson, and
  R.~Greenstadt, ``Git blame who?: Stylistic authorship attribution of small,
  incomplete source code fragments,'' \emph{Proceedings on Privacy Enhancing
  Technologies}, vol. 2019, pp. 389--408, 07 2019.

\bibitem{ekoko08}
E.~Duala-Ekoko and M.~P. Robillard, ``Tracking code clones in evolving
  software,'' in \emph{29th International Conference on Software Engineering
  (ICSE'07)}, 2007, pp. 158--167.

\bibitem{fowler99}
\BIBentryALTinterwordspacing
M.~Fowler, \emph{Refactoring - Improving the Design of Existing Code}, ser.
  Addison Wesley object technology series.\hskip 1em plus 0.5em minus
  0.4em\relax Addison-Wesley, 1999. [Online]. Available:
  \url{http://martinfowler.com/books/refactoring.html}
\BIBentrySTDinterwordspacing

\bibitem{geiger06}
R.~Geiger, B.~Fluri, H.~C. Gall, and M.~Pinzger, ``Relation of code clones and
  change couplings,'' in \emph{Fundamental Approaches to Software Engineering},
  L.~Baresi and R.~Heckel, Eds.\hskip 1em plus 0.5em minus 0.4em\relax Berlin,
  Heidelberg: Springer Berlin Heidelberg, 2006, pp. 411--425.

\bibitem{gitblame22}
\BIBentryALTinterwordspacing
{git-scm.com}, ``Show what revision and author last modified each line of a
  file,'' \emph{git--blame Manual Version 2.40.0}, 06 2022. [Online].
  Available: \url{https://git-scm.com/docs/git-blame}
\BIBentrySTDinterwordspacing

\bibitem{goon17}
A.~Goon, Y.~Wu, M.~Matsushita, and K.~Inoue, ``Evolution of code clone ratios
  throughout development history of open-source c and c++ programs,''
  \emph{International Workshop on Software Clones (IWSC2017)}, pp. 47--50,
  2017.

\bibitem{harder12}
J.~Harder, ``Code clone authorship — a first look,'' \emph{Softwaretechnik
  Trends}, vol.~32, no.~2, pp. 25--26, 2012.

\bibitem{honda14}
A.~Honda, H.~Aman, T.~Sasaki, and M.~Kawahira, ``Investigation of convergence
  tendency of code clone ratio in open source development,'' \emph{IEICE
  Transactions D}, vol. J97-D, no.~7, pp. 1213--1215, 2014.

\bibitem{hu14}
H.~Hu, H.~Zhang, J.~Xuan, and W.~Sun, ``Effective bug triage based on
  historical bug-fix information,'' in \emph{2014 IEEE 25th International
  Symposium on Software Reliability Engineering}, 2014, pp. 122--132.

\bibitem{inoue21-2}
K.~Inoue, ``Introduction to code clone analysis,'' in \emph{Code Clone
  Analysis}, K.~Inoue and C.~K. Roy, Eds.\hskip 1em plus 0.5em minus
  0.4em\relax Springer Singapore, 2021, pp. 3--27.

\bibitem{inoue21}
K.~{Inoue} and C.~K. {Roy}, Eds., \emph{Code Clone Analysis, Research, Tools
  and Practices}.\hskip 1em plus 0.5em minus 0.4em\relax Springer, 2021.

\bibitem{kalgutkar19}
\BIBentryALTinterwordspacing
V.~Kalgutkar, R.~Kaur, H.~Gonzalez, N.~Stakhanova, and A.~Matyukhina, ``Code
  authorship attribution: Methods and challenges,'' \emph{ACM Comput. Surv.},
  vol.~52, no.~1, feb 2019. [Online]. Available:
  \url{https://doi.org/10.1145/3292577}
\BIBentrySTDinterwordspacing

\bibitem{kamiya21}
T.~Kamiya, ``Ccfinderx: An interactive code clone analysis environment,'' in
  \emph{Code Clone Analysis}, K.~Inoue and C.~K. Roy, Eds.\hskip 1em plus 0.5em
  minus 0.4em\relax Springer Singapore, 2021, pp. 31--44.

\bibitem{Kamiya2002CCFinder}
T.~Kamiya, S.~Kusumoto, and K.~Inoue, ``Ccfinder: A multilinguistic token-based
  code clone detection system for large scale source code,'' \emph{IEEE Trans.
  Software Eng.}, vol.~28, pp. 654--670, 2002.

\bibitem{kim05}
M.~Kim, V.~Sazawal, D.~Notkin, and G.~Murphy, ``An empirical study of code
  clone genealogies,'' \emph{ACM SIGSOFT Software Engineering Notes}, vol.~30,
  no.~5, pp. 187--196, 2005.

\bibitem{koschke07}
R.~Koschke, ``Survey of research on software clones,'' in \emph{Dagstuhl
  Seminar Proceedings}.\hskip 1em plus 0.5em minus 0.4em\relax Schloss
  Dagstuhl-Leibniz-Zentrum f{\"u}r Informatik, 2007.

\bibitem{lee22}
\BIBentryALTinterwordspacing
Z.~Li, G.~Q. Chen, C.~Chen, Y.~Zou, and S.~Xu, ``Ropgen: Towards robust code
  authorship attribution via automatic coding style transformation,'' in
  \emph{Proceedings of the 44th International Conference on Software
  Engineering}, ser. ICSE '22.\hskip 1em plus 0.5em minus 0.4em\relax New York,
  NY, USA: Association for Computing Machinery, 2022, p. 1906–1918. [Online].
  Available: \url{https://doi.org/10.1145/3510003.3510181}
\BIBentrySTDinterwordspacing

\bibitem{lin14}
Y.~Lin, Z.~Xing, X.~Peng, Y.~Liu, J.~Sun, W.~Zhao, and J.~Dong, ``Clonepedia:
  Summarizing code clones by common syntactic context for software
  maintenance,'' in \emph{2014 IEEE International Conference on Software
  Maintenance and Evolution}.\hskip 1em plus 0.5em minus 0.4em\relax IEEE,
  2014, pp. 341--350.

\bibitem{linares12}
M.~Linares-Vásquez, K.~Hossen, H.~Dang, H.~Kagdi, M.~Gethers, and
  D.~Poshyvanyk, ``Triaging incoming change requests: Bug or commit history, or
  code authorship?'' in \emph{2012 28th IEEE International Conference on
  Software Maintenance (ICSM)}, 2012, pp. 451--460.

\bibitem{mondal21}
M.~Mondal, C.~K. Roy, B.~Roy, and K.~A. Schneider, ``Fleccs: A technique for
  suggesting fragment-level similar co-change candidates,'' in \emph{2021
  IEEE/ACM 29th International Conference on Program Comprehension (ICPC)},
  2021, pp. 160--171.

\bibitem{moriwaki14}
T.~Moriwaki, H.~Igaki, Y.~Yamanaka, N.~Yoshida, S.~Kusumoto, and K.~Inoue,
  ``Towards an analysis of who creates clone and who reuses it,'' in
  \emph{Proceedings of the Eighth International Workshop on Software Clones
  (IWSC 2014), Antwerp, Belgium}, 2014, pp. 1--5.

\bibitem{rabbani22}
S.~M. Rabbani, N.~Ahmad~Gulzar, S.~Arshad, S.~Abid, and S.~Shamail, ``A
  comparative analysis of clone detection techniques on semanticclonebench,''
  in \emph{2022 IEEE 16th International Workshop on Software Clones (IWSC)},
  2022, pp. 16--22.

\bibitem{rattan13}
D.~Rattan, R.~Bhatia, and M.~Singh, ``Software clone detection: A systematic
  review,'' \emph{Information and Software Technology}, vol.~55, no.~7, pp.
  1165--1199, 2013.

\bibitem{roy09}
C.~K. Roy, J.~R. Cordy, and R.~Koschke, ``Comparison and evaluation of code
  clone detection techniques and tools: A qualitative approach,'' \emph{Science
  of Computer Programming}, vol.~74, no.~7, pp. 470 -- 495, 2009.

\bibitem{roy07}
C.~K. Roy and J.~R. Cordy, ``A survey on software clone detection research,''
  \emph{Queen’s School of Computing TR}, vol. 541, no. 115, pp. 64--68, 2007.

\bibitem{OREO-FSE2018}
V.~Saini, F.~Farmahinifarahani, Y.~Lu, P.~Baldi, and C.~V. Lopes, ``Oreo:
  Detection of clones in the twilight zone,'' in \emph{Proceedings of the 2018
  {ACM} Joint Meeting on European Software Engineering Conference and Symposium
  on the Foundations of Software Engineering, {ESEC/SIGSOFT} {FSE} 2018, Lake
  Buena Vista, FL, USA, November 04-09, 2018}, 2018, pp. 354--365.

\bibitem{Sajnani2016SourcererCC}
\BIBentryALTinterwordspacing
H.~Sajnani, V.~Saini, J.~Svajlenko, C.~K. Roy, and C.~V. Lopes, ``Sourcerercc:
  Scaling code clone detection to big-code,'' in \emph{Proceedings of the 38th
  International Conference on Software Engineering}, ser. ICSE '16.\hskip 1em
  plus 0.5em minus 0.4em\relax New York, NY, USA: ACM, 2016, pp. 1157--1168.
  [Online]. Available: \url{http://doi.acm.org/10.1145/2884781.2884877}
\BIBentrySTDinterwordspacing

\bibitem{semura17}
Y.~Semura, N.~Yoshida, E.~Choi, and K.~Inoue, ``Ccfindersw: Clone detection
  tool with flexible multilingual tokenization,'' in \emph{Proceedings of the
  24th Asia-Pacific Software Engineering Conference (APSEC 2017), Nanjing,
  China}, 2017, pp. 654--659.

\bibitem{tokui20}
S.~Tokui, N.~Yoshida, E.~Choi, and K.~Inoue, ``Clone notifier: Developing and
  improving the system to notify changes of code clones,'' in \emph{2020 IEEE
  27th International Conference on Software Analysis, Evolution and
  Reengineering (SANER)}, 2020, pp. 642--646.

\bibitem{walker19}
\BIBentryALTinterwordspacing
A.~Walker, T.~Cerny, and E.~Song, ``Open-source tools and benchmarks for
  code-clone detection: Past, present, and future trends,'' \emph{SIGAPP Appl.
  Comput. Rev.}, vol.~19, no.~4, p. 28–39, jan 2020. [Online]. Available:
  \url{https://doi.org/10.1145/3381307.3381310}
\BIBentrySTDinterwordspacing

\bibitem{Zhao2018DeepSim}
G.~Zhao and J.~Huang, ``Deepsim: Deep learning code functional similarity,'' in
  \emph{Proceedings of the 2018 26th ACM Joint Meeting on European Software
  Engineering Conference and Symposium on the Foundations of Software
  Engineering}, ser. ESEC/FSE 2018.\hskip 1em plus 0.5em minus 0.4em\relax New
  York, NY, USA: ACM, 2018, pp. 141--151.

\end{thebibliography}

\end{document}